\documentclass[twocolumn,prb]{revtex4-1}
\usepackage{graphicx}

\usepackage{amssymb}
\usepackage{amsmath}
\usepackage[latin1]{inputenc}
\usepackage{bm}

\newcommand{\xib}{\boldsymbol{\xi}}

\begin{document}

\title{Quantum pumping in graphene}
\author{E. Prada, P. San-Jose, and H. Schomerus}
\affiliation{Department of Physics, Lancaster University,
Lancaster,  LA1 4YB, United Kingdom}
\date{\today}
\pacs{}

\begin{abstract}
We show that graphene-based quantum pumps can tap into evanescent
modes, which penetrate deeply into the device as a consequence of
Klein tunneling. The evanescent modes dominate pumping at the
Dirac point, and give rise to a universal response under weak
driving for short and wide pumps, in close analogy to their role
for the minimal conductivity in ballistic transport. In contrast,
evanescent modes contribute negligibly to normal pumps. Our
findings add a new incentive for the exploration of graphene-based
nanoelectronic devices.
\end{abstract}

\maketitle

\section{Introduction} Quantum pumps transfer electrons between two
reservoirs by externally varying their scattering properties over time. This
concept has attracted much attention since its inception, due in part to its
promise for practical applications in nanoelectronics \cite{Switkes:S99,
Watson:PRL03} and for the definition of a current standard,\cite{Wright:09,
Pekola:NP07, Kouwenhoven:PRL91} but also because of its elegant theoretical
description in terms of the geometry of the control parameter
manifold.\cite{Buttiker:ZPB94, Brouwer:PRB98, Makhlin:PRL01, Aharonov:PRL87}

Efficient quantum pumping requires strong but energy-dependent coupling of
the pump to the reservoirs. In normal systems, pumping is therefore
constrained to propagating modes, while the poorly coupled evanescent modes
decay rapidly away from the contacts and therefore cannot contribute to the
pumped charge. Here, we show that the discovery of graphene-based materials
\cite{Novoselov:S04} calls for a revision of such common concepts in quantum
pumping. In graphene, the low-energy charge carriers are described by a
massless Dirac equation,\cite{Novoselov:N05} and the unique feature of
chirality suppresses backscattering at interfaces, resulting in the so-called
Klein paradox by which charge carriers are difficult to confine.%
\cite{Neto:RMP09, Katsnelson:NP06, Beenakker:RMP08} This seems to inhibit the
prospects of quantum pumping---unless one properly accounts for the effects
of chirality on the evanescent modes. These effects so far have been explored
only for stationary ballistic transport, where evanescent chiral electrons
manifest themselves in macroscopic quantum tunneling close to the Dirac point
of charge-neutral graphene.\cite{Katsnelson:EPJB06, Tworzydlo:PRL06} We show
that the scattering of these evanescent modes is sufficiently
energy-dependent so that they contribute significantly to quantum pumping.
Close to the Dirac point, they deliver the dominant contribution to the
pumped charge, which can be characterized by a universal dimensionless
pumping efficiency.

We start this paper in Sec.\ \ref{sec:2}  with a description of the
employed model (a quantum pump driven by two electrostatic gates) and  method
(the scattering approach to adiabatic quantum pumping). Our results are
presented in Sec.\ \ref{sec:3}. The concluding remarks in Sec.\ \ref{sec:4}
also contain a discussion of practical advantages of graphene-based quantum
pumps.

\begin{figure}[top]
\includegraphics[width=\linewidth]{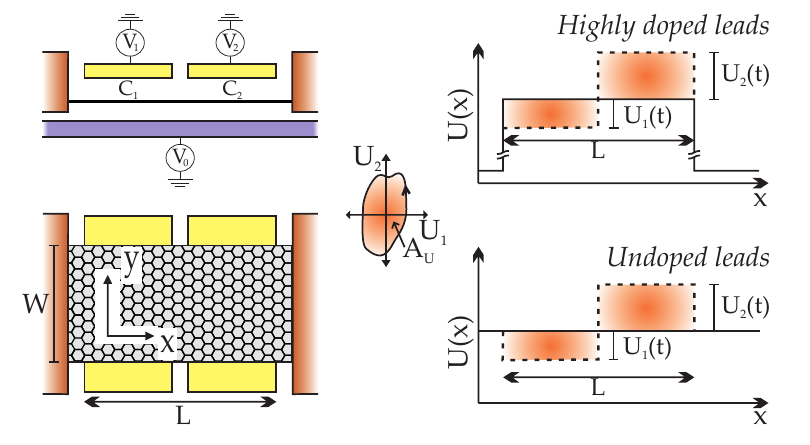}
\caption{\label{fig:system} (Color online) Left panels: Illustration of a
graphene quantum pump with highly doped (metallic) contacts. Two gates at voltage $V_1(t)$ and
$V_2(t)$ control the time-dependent onsite energy $U_1(t)$ and
$U_2(t)$ in the graphene flake over a pumping cycle.
This induces a charge transport between two contact electrodes, separated by
a distance $L$. The working point of the device can be further controlled
by an additional back gate with voltage $V_0$. Right panels: Snapshots of the onsite potential
along the quantum pump, for highly doped (metallic) contacts (top) and undoped contacts (bottom).
The solid base line shows the potential at the working point.
We show that evanescent modes induced by metallic contacts enable finite
charge pumping at vanishing nominal charge-carrier density, in striking
contrast to the case of normal pumps where the nanoribbon is replaced by a
normal conductor.}
\end{figure}

\section{\label{sec:2} Model and method}

In order to characterize the unique features of quantum pumping in
graphene, we compare the behavior of four different setups: graphene pumps
with either heavily doped contacts or undoped contacts are contrasted with
normal-conductor pumps with these two types of contacts. In the absence of
driving (i.e., at the working point of the pump), the systems with undoped
contacts have a uniform carrier density, equal in both the leads and in the
central pumping region. In contrast, the charge carrier density in heavily
doped contacts is very large; this can be realized by metallic contacts
\cite{Schomerus:PRB07}. We assume that the charge carrier density in the
system can be further uniformly shifted by a back gate covering both the
pumping region and leads, thereby tuning the working point of the device. For
heavily doped contacts, the back-gate-induced change of the charge carrier
density in the leads can be neglected. Since only the setup with highly doped
contacts induces evanescent modes, this comparison allows to isolate the
requirements for the successful deployment of such modes in quantum pumping.

In all of these four cases, we consider a standard pumping mechanism,
which is shown for the example of a graphene pump with heavily doped
(metallic) contacts in the left panels of Fig.\ \ref{fig:system}.  The pump
is operated by a cyclic raising and lowering of the potential in two
additional electrostatic gates, which modulate the charge carrier density in
the two halves of the system due to their effect on the onsite potential $U$.
This induces charge transfer between the two electronic reservoirs of the
contacts. The bottom left panel shows a rotated view of the pump, which is
further characterized by the width $W$ and length $L$ of the pumping region.
The top right panel shows a snapshot of the onsite potential (dashed line)
for the case of highly n-doped contacts (the solid line is the onsite
potential at the working point). The bottom right panel shows a corresponding
snapshot for undoped leads.

An elegant formulation of quantum pumping is afforded by the scattering
approach \cite{Brouwer:PRB98, Makhlin:PRL01}, which considers the time
dependence of the scattering matrix $S(t)$. In the minimal case of adiabatic
driving with two independent parameters $\xib=\{\xi_1, \xi_2\}$ and single
channel reservoirs, the charge transferred across the non-interacting
scattering region reduces to an integral over the area enclosed by the
driving path in two-dimensional (2D) parameter space \footnote{Throughout
this work, the term \emph{charge} refers to the \emph{number} of electrons.},
\begin{subequations}\label{Q}%
\begin{eqnarray}
Q&=&\int d\xi_1 d\xi_2\,\, \partial^2_{\xi}Q(\xib),\\
\partial^2_{\xi}Q
&\equiv&\frac{1}{2\pi}\left(\frac{\partial
T}{\partial\xi_1}\frac{\partial
\phi}{\partial\xi_2}-\frac{\partial
T}{\partial\xi_2}\frac{\partial \phi}{\partial\xi_1}\right).
\end{eqnarray}
\end{subequations}
The transmission probability
$T$ and the scattering phase $\phi=\alpha-\beta$ above are
determined by the scattering matrix
\[
S=
\left(\begin{array}{cc}
r & t'\\
t& r'
\end{array}\right)
= e^{i\gamma}\left(\begin{array}{cc}
\sqrt{1-T}e^{i\alpha} & -\sqrt{T}e^{i\beta}\\
\sqrt{T}e^{-i\beta} & \sqrt{1-T}e^{-i\alpha}
\end{array}\right),
\]
where $r$ ($r'$) and $t$ ($t'$) are reflection and transmission
amplitudes for electrons arriving from the left (right) reservoir.

In a quasi one-dimensional setup with more than one channel, indexed by
quantum number $q$, the total pumped charge will be a sum over channels
$Q=\sum_{q}Q_{q}$ as long as they can be considered independent. This is the
case for the quantum pump setup depicted in Fig.\ \ref{fig:system}. Across
the width $W$ of the system, the onsite potential $U$ is $y$ independent, so
that different scattering channels $q$ remain decoupled. We model this
profile by two abrupt potential steps in the $x$ direction of equal length
$L/2$ and assume the two driving parameters $\xib=\{U_1(t),U_2(t)\}$ have
zero average, and maximum amplitudes $\delta U_1$ and $\delta U_2$,
respectively. A back gate is used to control the carrier concentration in
both pump and leads, which we parameterize by the average Fermi momentum
$k_F$ in the pumping region. Undoped contacts have the same Fermi
momentum $k_F$ as the pumping region, while heavily doped contacts have a
much larger Fermi momentum, which can effectively be taken as infinite.
\footnote{All our results therefore do not feature an
additional Fermi momentum for the leads.} By employing the Dirac equation
with negligible inter-valley scattering for graphene \cite{Neto:RMP09}, and
an effective mass approximation for the normal system, this model allows us
to compute the transmission probabilities and phases $T_{q}$ and $\phi_{q}$
by simple wave matching.

\section{\label{sec:3} Results and discussion} The finite length $L$ of the pump provides a
natural scale for the scattering problem. It fixes the energy scales
$E_L^G=\hbar v_0/L$ (with Dirac velocity $v_0$) in the graphene case, and
$E_L^N=\hbar^2/(2m^*L^2)$  (with effective mass $m^*$) for the normal
conductor, which are related to the level spacing of the isolated pump. These
energies determine two possible driving regimes, depending on the maximum
amplitude $\delta u_i$ of the dimensionless driving energies $u_i\equiv
U_i/E_L$. In the weak driving regime, $\delta u_i\ll 1$, the charge pumped in
channel $q$ can be approximated by
\begin{equation}\label{weakdriving}
  Q_{q}\approx\partial_{u}^2 Q_{q}(0)\int d u_1 d u_2=A_{u} \partial_{u}^2 Q_{q}(0),
\end{equation}
where $A_{u}\sim \delta u_1\delta u_2$ is the small area enclosed
in parameter space by the driving cycle, wherein $\partial_{u}^2
Q_{q}(u_1,u_2)$ can be approximated by a constant.
Away from this regime the integral in Eq.\ (\ref{Q}) has to be
performed numerically.

\begin{figure}[top]
\includegraphics[width=\linewidth]{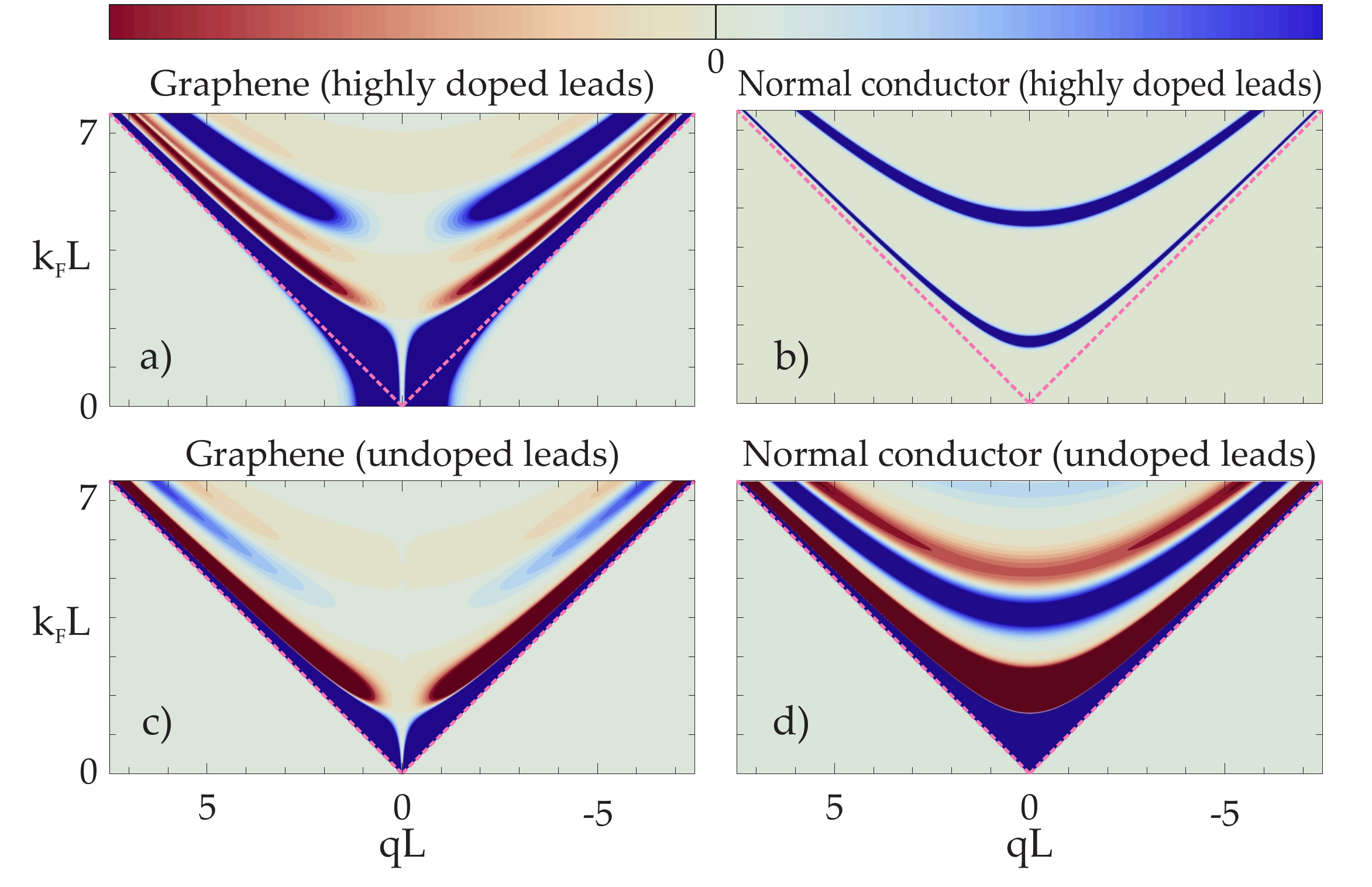}
\caption{\label{fig:chiq}(Color online) Momentum distribution of
pumped charge per mode $\chi^u_q$ as a function of mode index $q$
for varying carrier concentration (parameterized by the Fermi momentum $k_F$).
Blue and brown represent opposite directions of pumping (left to
right or right to left).  In graphene, the propagating mode with
$q=0$ (normal incidence) cannot be pumped due to the Klein
paradox. In the case of graphene with heavily doped leads, however, significant pumping
is possible due to the contribution of the evanescent modes
($|q|>k_F$, delineated by the dashed line), which dominate around
the Dirac point ($k_F=0$). The other pumps can only drive current
through the propagating modes.}
\end{figure}

For the graphene pump with heavily doped contacts, Eq.\ (\ref{weakdriving})
yields
\begin{eqnarray}\label{QGE}
   Q^{\mathrm{gr-\infty}}_q&=&\pm A_u\frac{k_FL}{\pi}\frac{(qL)^2}{k_xL}\\
   &\times&\frac{\sin^2(k_xL) \left[\sin (2k_xL)-2k_xL\cos (2k_xL)\right]}{\left[(k_xL)^2+(qL)^2\sin^2 (2k_xL)\right]^2}.\nonumber
\end{eqnarray}
Here the $\pm$ sign denotes whether the pump is doped with electrons (plus)
or holes, and $k_x=\sqrt{k_F^2-q^2}$ is the electron's momentum along the
transport direction in the pumping region. Momentum $k_x$ is real for
propagating modes $|q|<k_F$ and imaginary for modes $|q|>k_F$ that are
evanescent in the pump. In contrast, a graphene pump with undoped leads has
no incoming lead modes that become evanescent in the pump (in the limit
$\delta u_i\ll 1$), since the Fermi momentum in the pump at the
working point is identical to the Fermi momentum in the contacts. A graphene
setup with undoped leads pumps a charge
\begin{equation} Q^{\mathrm{gr-0}}_q=A_u\frac{k_FL}{\pi}\frac{2(qL)^2\cos^2(k_xL)\sin^3(k_xL)}{(k_xL)^4},
\end{equation}
where $k_x$ is constrained to real values since $|q|\leq k_F$.

In both cases the pumped charge has a prefactor $A_u k_F$,
indicating that pumping is proportional to the pump's number
$N_p=4 k_F W/\pi$ of propagating electrons at the Fermi energy and
the dimensionless driving strength $A_u$. By factoring out these
two quantities, we obtain the dimensionless pumping response
\begin{equation}\label{chi}
  \chi_q^u\equiv\frac{\partial_{u}^2 Q_{q}}{N_p}\approx\frac{Q_q}{A_u
  N_p},
\end{equation}
which depends only on the system's scattering characteristics at a
given energy. Summing over all incoming modes gives the fraction
of (propagating) electrons in the pump that are pumped per cycle
and per unit dimensionless driving strength $A_u$,
\begin{equation}\label{chisum}
  \chi^u=\sum_q \chi_q^u=\frac{1}{N_p}\sum_q \partial_{u}^2 Q_q.
\end{equation} This defines the
total dimensionless pumping response per mode. For short and wide
pumps ($W\gg L$), the sum over modes can be approximated by an
integral.

Similar results can be derived for the normal pump,
\begin{eqnarray} \label{eq:normalpumpsq}
	Q^{\mathrm{n}}_q&=&A_u 8\frac{k_F^2K_x^3}{k_xL^2}\sin^2(k_xL)\\
	&\times& \frac{2k_xL(k_x^2-K_x^2)\cos(2k_xL)+(k_x^2+K_x^2)\sin(2k_xL)}{\left[4 k_x^2K_x^2\cos^2(2k_xL)+(k_x^2+K_x^2)^2\sin^2(2k_xL)\right]^2}\nonumber
\end{eqnarray}
Both the heavily doped contact ($Q^{\mathrm{n-\infty}}_q$) and undoped
contact  ($Q^{\mathrm{n-0}}_q$) normal cases can be derived from the above by
taking the $x$ momentum in the leads $K_x$ to infinity or $k_x$ respectively.

In Fig.\ \ref{fig:chiq} we  represent the pumping response $\chi^u_q$ of
Eq.~(\ref{chi}) as a function of $q$ and $k_F$ in the two graphene setups,
and compare these with the results of the normal pumps. For graphene we only
show results for electron carriers. If the leads are heavily doped the pumped
current reverts sign when the carriers in the central pump region are changed
from electrons to holes [cf.\ Eq.\ (\ref{QGE})]. In the case of undoped
leads, the pumped current remains the same and only reverts sign if one also
reverts the pumping cycle.

In all four panels, the dashed line delineates the threshold between
evanescent and propagating modes. The most evident feature in the graphene
pump with heavily doped leads [Fig.\ \ref{fig:chiq}(a)] is the contribution
of evanescent modes to pumping close to the Dirac point. This contribution is
absent in the other three cases, in which only propagating electrons are
pumped. The reason for this is the high transmission of evanescent modes in
graphene, which can be attributed to chirality and the Klein paradox
\cite{Katsnelson:NP06}. Unlike in the normal pump with heavily doped leads,
chirality conservation at the contact enables evanescent electrons to
populate the graphene pumping region for modes within a window of width
$\Delta q\sim 1/L$ around $q=0$. These evanescent modes contribute to pumping
because they are sensitive to the onsite potentials $U_i$ and have a finite
amplitude at both contacts, so that charge transfer between them is possible
over a pumping cycle. The finite contribution of the evanescent modes in
graphene is in striking contrast to the vanishing contribution of electrons
in the propagating mode $q=0$ (normal incidence), for which the transmission
$T_{q=0}=1$ is perfect at all energies because of the Klein paradox---these
modes are therefore insensitive to driving and cannot be pumped.

In a normal pump with heavily doped contacts [Fig.\ \ref{fig:chiq}(b)], the
large Fermi velocity mismatch suppresses the transparency of the contacts for
all modes except those at resonant tunneling. This confinement creates
sharply defined energy levels in the pump which are the origin of narrow
regions of finite pumping, and results in a threshold $k_F L\simeq 1$ below
which no pumping occurs. In particular, the contribution of evanescent modes
to pumping is strongly suppressed (no contribution outside of the dashed
line). Moreover, normal pumping in the limit of large lead doping is
directed, meaning that for a given orientation of the driving cycle, the
pumped current has the same sign for all energies.

For graphene and normal pumps with undoped leads [Fig.\ \ref{fig:chiq}(c) and
(d)], all incoming modes remain propagating in the pumping region. The main
difference between graphene and normal pumps in this limit is the effect of
the Klein paradox in the former, which suppresses pumping at $q=0$, just as
in the case of graphene with doped leads. Both pumps are open, and
consequently there is no energy threshold for pumping. The sign of the pumped
current is energy dependent, which is a generic feature of open pumps
(including graphene pump with doped leads).

\begin{figure}[top]
\includegraphics[width=0.9\linewidth]{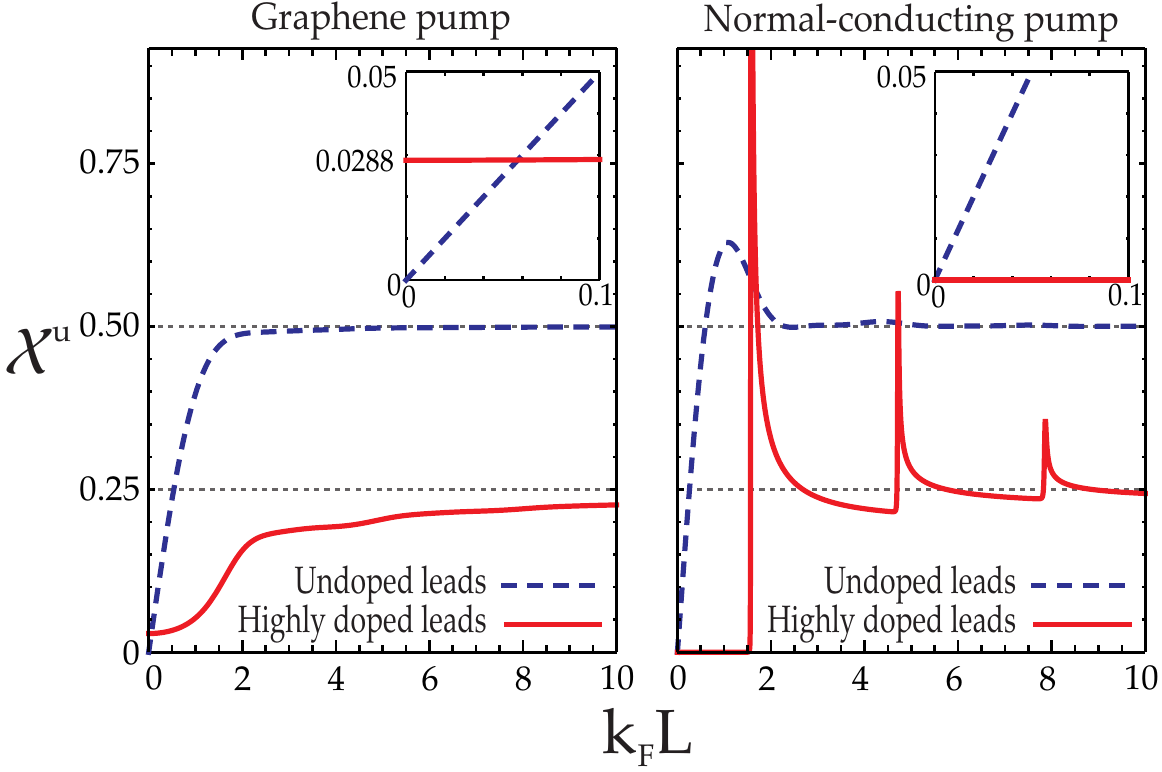}
\caption{\label{fig:chi} (Color online) Total dimensionless
response per mode $\chi^u$  for short and wide pumps ($W\gg L$),
as a function of the carrier concentration (parameterized by the Fermi momentum
$k_F$). Left panel: graphene pumps. Right panel: Normal-conducting pumps.
Solid lines correspond to pumps with highly doped leads (which tend to $1/4$
at $k_FL\gg1 $), while dashed lines correspond to pumps with undoped leads
(which tend to $1/2$).  At the charge neutrality ($k_FL\to 0$, see inset),
evanescent modes allow for a finite charge transfer in the
highly-doped-leads graphene pump, whose response saturates at a
universal (sample-independent) value of $0.0288$.}
\end{figure}

In Fig.\ \ref{fig:chi} we plot the total dimensionless pumping response
$\chi^u(k_F)$ of Eq. (\ref{chisum}) for the four types of pumps considered
(left panel: graphene pumps; right panel: normal-conducting pumps).
The regime of evanescent electron pumping in graphene with highly doped leads
is visible as a saturation at $k_F=0$ (see the solid curve in the inset). For
wide and short pumps ($W\gg L$), this saturation value is sample independent
and takes the dimensionless value
\begin{equation}
  \int_0^\infty dq \frac{\sinh^2(q)\left[2q\cosh(2q)-\sinh(2q)\right]}{\pi q^3 \cosh^4(2q)}
  =0.0288,
\end{equation}
which is the analogue of the minimal conductivity in the context of pumping
\cite{Katsnelson:EPJB06,Tworzydlo:PRL06}. In contrast, all other pumps have a
vanishing pumping response at depletion. At energies $k_F\gtrsim 1/L$, the
pumping response rises to $1/2$ and $1/4$ in the cases of undoped and highly
doped leads, respectively. The  normal pump with highly doped leads, however,
only operates above a finite carrier-concentration threshold corresponding to
the position of the first resonant tunneling subband as mentioned above.

\section{\label{sec:4} Concluding remarks} In summary, we find that evanescent modes can
contribute significantly to graphene-based quantum pumping, to the extent
that they provide the dominant contribution to the pumped charge when the
system is operated close to the charge-neutrality point. Our comparison to
results for normal pumps reveals that this effect is intimately related to
chirality and the Klein paradox, and therefore arises from the unique
low-energy properties of charge carriers in graphene. For the case of short
and wide pumps, the evanescent pumping regime is characterized by a
sample-independent universal value of the dimensionless pumping response.

Practical considerations  point towards additional advantages of
graphene-based quantum pumps. Firstly, in a realistic experimental setup, the
principal pump driving parameters are not the onsite energies $U_i$, but gate
voltages $V_i$ (see Fig. \ref{fig:system}) which manipulate the locally
induced charge density $n_i(t)=V_i(t) C_i$ under each gate depending on the
capacitances $C_i$. The onsite energies then follow from the difference of
the Fermi energy and the local position of the Dirac point. Neglecting
details of the screening, for graphene $n_i=W u_i^2/(\pi L)$, whilst for the
normal case $n_i=W u_i/(2\pi L)$ (in the case of graphene with highly doped
leads, a precise modeling would also have to take care of  charge carriers
populating the evanescent modes). In terms of these charge densities, the
relevant response function is
\begin{equation}
  \chi^n=\chi^u\det\left(\frac{\partial u_i}{\partial n_j}\right),
\end{equation}
involving the Jacobian between the $u$ and the $n$ variables. Due
to the divergence $\partial u_i/\partial n_i\propto n_i^{-1/2}$
close to the Dirac point, the experimental pumping response
$\chi^n$  for graphene pumps is expected to rapidly increase as
one approaches charge neutrality, while it vanishes for normal
pumps. Secondly, graphene-based quantum pumping promises to
display an enhanced robustness against thermal effects. Thermal
smearing of the pumping response \cite{Vavilov:PRB01} occurs at
temperatures of order $E_L/k_B$, which are considerably higher in
graphene than in normal pumps (this is also one of the reasons for
the temperature robustness of other transport effects in graphene
\cite{Novoselov:NP06}). For the same reason, mechanisms limiting
the pumping frequency are expected to be less stringent in
graphene than in normal pumps. These considerations, together with
the long coherence length and high mobility in graphene, lead us
to believe that graphene-based quantum pumps have good chances to
become an experimental reality.

We acknowledge support from the European Commission, Marie Curie
Excellence Grant MEXT-CT-2005-023778.

\bibliography{biblio}

\end{document}